\documentclass[journal]{IEEEtran}

\usepackage{cite}          
\usepackage{amsmath}       
\usepackage{amsfonts}      
\usepackage{amssymb}       

\usepackage{array}         
\usepackage{url}           
\usepackage{booktabs}      
\usepackage{multirow}      
\usepackage{multicol}      
\usepackage{color}         
\usepackage{xcolor}        
\usepackage{textcomp}      
\usepackage{siunitx}       
\usepackage{xspace}

\ifCLASSINFOpdf
  \usepackage[pdftex]{graphicx}
  \graphicspath{{./figures/}{./images/}{./plots/}}
  \DeclareGraphicsExtensions{.pdf,.jpeg,.jpg,.png}
\else
  \usepackage[dvips]{graphicx}
  \graphicspath{{./figures/}{./images/}{./plots/}}
  \DeclareGraphicsExtensions{.eps}
\fi

\ifCLASSOPTIONcompsoc
  \usepackage[caption=false,font=normalsize,labelfont=sf,textfont=sf]{subfig}
\else
  \usepackage[caption=false,font=footnotesize]{subfig}
\fi

\usepackage{float}         

\usepackage{bm}            
\usepackage{cases}         
\usepackage{theorem}       
\usepackage{algorithm}     
\usepackage{algorithmicx}  
\usepackage{algpseudocode} 

\usepackage{tabularx}      
\usepackage{longtable}     
\usepackage{rotating}      

\usepackage{enumerate}     
\usepackage{enumitem}      
\usepackage{balance}       
\usepackage{flushend}      
\usepackage{lipsum}        
\usepackage{blindtext}     

\usepackage{booktabs}
\usepackage{siunitx}
\newcolumntype{Y}{>{\raggedright\arraybackslash}X}

\usepackage{booktabs,tabularx,siunitx}
\begin{document}

\title{MCR-VQGAN: A Scalable and Cost-Effective Tau PET Synthesis Approach for Alzheimer's Disease Imaging}

\author{Jin~Young~Kim,
        Jeremy~Hudson,
        Jeongchul~Kim,
        Qing~Lyu*,
        and~Christopher~T.~Whitlow*%

\thanks{Jin~Young~Kim is with the Department of Biomedical Engineering, Wake Forest University School of Medicine, Winston-Salem, NC 27101, United~States. E-mail: jin.kim@wfusm.edu}%
\thanks{Jeongchul~Kim is with the Department of Radiology, Wake Forest School of Medicine, Winston-Salem, NC 27101, United~States. E-mail: jeongchul.kim@wfusm.edu}%
\thanks{Christopher~T.~Whitlow, Qing~Lyu, and Jeremy~Hudson are with the Department of Radiology and Biomedical Imaging, Yale School of Medicine, New Haven, CT 06510 USA.}%
\thanks{(\textit{Corresponding authors: Qing Lyu; Christopher T. Whitlow.}) E-mail: qing.lyu@yale.edu; christopher.whitlow@yale.edu}}

\markboth{Journal Name, Vol.~XX, No.~X, Month~Year}%
{Author \MakeLowercase{\textit{et al.}}: Short Title for Header}



\maketitle


\begin{abstract}
Tau positron emission tomography (PET) is a critical diagnostic modality for Alzheimer's disease (AD), but its widespread clinical adoption is hindered by radiation exposure, limited availability, high clinical workload, and substantial financial costs. To address these limitations, we propose the Multi-scale CBAM Residual Vector Quantized Generative Adversarial Network (MCR-VQGAN) to synthesize high-fidelity tau PET images from structural T1-weighted MRI. MCR-VQGAN advances the standard VQGAN architecture through three enhancements: multi-scale convolutions, ResNet blocks, and Convolutional Block Attention Modules (CBAM), which collectively improve the capture of local and global features. Using 222 paired T1-weighted MRI and tau PET scans from the ADNI database, we trained and compared MCR-VQGAN against cGAN, WGAN-GP, CycleGAN, and baseline VQGAN. MCR-VQGAN achieved superior image synthesis performance across all metrics (MSE $= 0.0056 \pm 0.0061$, PSNR $= 30.65 \pm 4.47$ dB, SSIM $= 0.9263 \pm 0.0469$). A CNN-based AD classifier trained on real tau PET achieved comparable accuracy on real ($63.64\%$) and synthetic ($65.91\%$) images, indicating that diagnostically relevant features are preserved. Regional SUVR-equivalent analysis across Braak-defined ROIs further indicated strong agreement between real and synthetic tau PET (Pearson $r = 0.78$--$0.88$; ICC $= 0.71$--$0.84$), with the strongest agreement in Braak V/VI (ICC $= 0.838$). Together, these results suggest that MCR-VQGAN offers a promising and scalable surrogate for conventional tau PET imaging, potentially improving the accessibility of tau biomarkers for AD research and clinical workflows.
\end{abstract}


\IEEEpeerreviewmaketitle


\section{Introduction}
\IEEEPARstart{A}{lzheimer's} disease (AD) is a prevalent progressive neurodegenerative disorder characterized by gradual declines in memory, cognitive function, and the ability to perform daily activities~\cite{kumar2024alzheimer}. Currently, AD affects approximately 7.2 million Americans, and this number is expected to increase as the population ages~\cite{rajan2021population,better2023alzheimer}. As there is no cure for AD, early and accurate diagnosis is essential to enable timely interventions that can slow disease progression and optimize patient outcomes~\cite{rasmussen2019alzheimer,alzheimer20252025}. 

The accumulation of hyperphosphorylated tau protein into neurofibrillary tangles (NFTs) is a critical AD biomarker, showing strong correlations with cognitive decline and neurodegeneration. Consequently, tau positron emission tomography (PET) has become a key imaging modality, enabling direct visualization and quantification of tau deposits~\cite{johnson2012brain,van2021imaging}. This facilitates accurate disease staging and progression monitoring, advancing AD diagnosis and research~\cite{johnson2012brain,frisoni2010clinical,van2021imaging}. However, its widespread clinical adoption is hindered by substantial financial cost, limited scanner availability, patient radiation exposure, and high clinical workload. These limitations restrict tau PET accessibility in both clinical and research settings~\cite{vermeiren2024survey}.

Recent advances in deep learning, including image-to-image translation and generation~\cite{pang2021image}, offer a potential solution to these challenges. Several studies have demonstrated the feasibility of generating synthetic PET images from more accessible modalities such as MRI using generative adversarial networks (GANs)~\cite{lan2021three,sikka2025mri,shin2020ganbert,wei2019predicting,pan2018synthesizing,zhang2022bpgan}. More recently, diffusion models (e.g., DDPMs) have achieved state-of-the-art (SOTA) performance in medical image synthesis~\cite{dorjsembe2022three,pinaya2022brain,muller2022diffusion,huy2023denoising}. However, their computational intensity, large data requirements, and long processing times make them unsuitable for resource-constrained clinical workflows. In contrast, GANs offer faster inference, lower computational demands, and strong performance with limited data. Thus, GANs remain the preferred choice for medical image synthesis, particularly in PET applications.

Among GAN-based architectures, Vector Quantized Generative Adversarial Network (VQGAN) has emerged as a promising approach, leveraging vector quantization to learn discrete latent representations while maintaining the benefits of adversarial training for high-fidelity image synthesis~\cite{esser2021taming}. While VQGAN improves on conventional GANs, its application to tau PET synthesis suffers from: (1) over-smoothing that blurs crucial details, (2) training instability/mode collapse, and (3) poor preservation of diagnostic features~\cite{sikka2025mri}. These challenges arise from architectural limitations in standard GANs, which often struggle to capture global context and model long-range spatial dependencies in brain imaging data.

To overcome these limitations, we propose Multi-scale CBAM Residual Vector Quantized Generative Adversarial Network (MCR-VQGAN), a novel architecture that enhances standard VQGAN~\cite{esser2021taming} by integrating multi-scale convolutions, ResNet blocks~\cite{he2016deep}, and Convolutional Block Attention Modules (CBAM)~\cite{woo2018cbam} to synthesize high-fidelity tau PET images from T1-weighted MRI. Specifically, we address the structural--functional gap in tau PET synthesis by
\begin{itemize}
    \item Expanding receptive fields through multi-scale convolutional layers to capture global brain structure and inter-regional dependencies.
    \item Improving feature propagation and reducing vanishing gradients through the use of ResNet blocks.
    \item Leveraging vector quantization to mitigate mode collapse and stabilize training.
    \item Preserving diagnostic features in anatomically relevant regions through CBAM integration.
\end{itemize}
Ultimately, this framework addresses the critical barriers to widespread tau PET imaging: significant financial costs, cumulative radiation exposure, increased clinical workload, and limited facilities. By providing a robust method for generating high-fidelity tau PET from MRI, this work offers an accessible, cost-effective, and non-invasive alternative to enhance both AD research and clinical workflows.

\begin{figure*}[!t]
\centering
\includegraphics[width=0.9\textwidth]{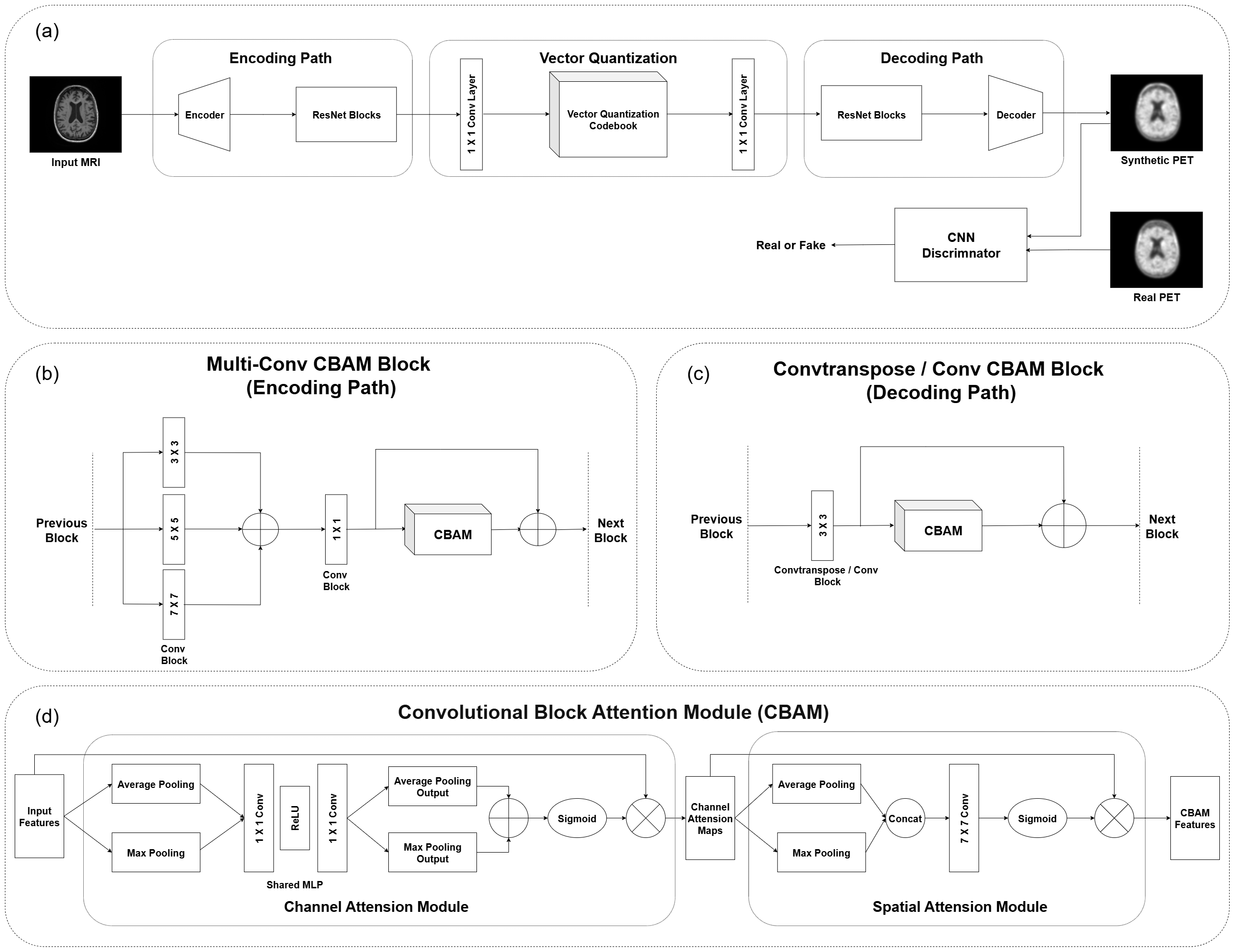}
\caption{Overview of the proposed MCR-VQGAN framework.
(a) The end-to-end synthesis pipeline, showing the \emph{encoding path} (encoder and ResNet blocks), the vector quantization module, the \emph{decoding path} (decoder and ResNet blocks), and the discriminator.
(b) The multi-scale convolutional block used in the encoding path.
(c) The standard convolutional block used in the decoding path.
(d) The Convolutional Block Attention Module (CBAM)~\cite{woo2018cbam} architecture.}
\label{fig:architecture}
\end{figure*}

\section{Related Work}
\subsection{GANs For Medical Image Synthesis}
A GAN is a generative model that generates realistic synthetic data through an adversarial process, training two competing networks (generator and discriminator) simultaneously. This competitive training enables GANs to approximate complex data distributions and synthesize high-fidelity data~\cite{goodfellow2020generative}. Consequently, GANs are widely employed for cross-modality medical image synthesis across various modalities, as they can capture complex nonlinear mappings between modalities and generate high-fidelity images with computational efficiency~\cite{pang2021image}. Among these applications, MRI-to-CT synthesis has been most studied due to the widespread clinical use of both modalities, which has produced large datasets and facilitated the development of robust generative models. 

Conditional GANs (cGANs)~\cite{isola2017image} have been widely applied in paired image-to-image translation tasks, particularly for MRI-to-CT synthesis, where they have yielded superior performance by learning direct pixel-to-pixel mappings~\cite{nie2018medical,emami2018generating,maspero2018dose,boni2020mr,ranjan2022gan,zhao2023ct}. Beyond MRI-to-CT translation, cGANs have been extended to intra-MRI translation (e.g., T1-to-T2~\cite{yang2018mri}, T1-to-FLAIR~\cite{yu20183d}) and multimodal synthesis such as T1/T2-to-MRA~\cite{olut2018generative}. However, their reliance on paired data limits applicability in medical imaging. To overcome this, Cycle-Consistent GAN (CycleGAN)~\cite{zhu2017unpaired} was introduced for unpaired translation through bidirectional cycle-consistency, enabling training without paired datasets. CycleGAN and its variants have demonstrated robust performance and strong generalizability in MRI-to-CT translation across heterogeneous datasets~\cite{wolterink2017deep,hiasa2018cross,yang2018unpaired}. It has also been successfully applied to intra-MRI translation tasks, such as T1-to-T2~\cite{welander2018generative,liu2019susan}. 

Further architectural innovations have been developed and extensively explored in medical image synthesis. Wasserstein GANs (WGANs)~\cite{arjovsky2017wasserstein}, for instance, have been leveraged for CT synthesis~\cite{kong2023synthetic,li2021low}, X-ray synthesis~\cite{senthil2024synthetic}, and retinal image generation~\cite{anaya2024wgan}, demonstrating improved training stability and image fidelity over conventional GANs. StyleGAN~\cite{karras2019style} has also gained attention for its style-based generator architecture, enabling fine-grained control over global structure and local texture. It has been utilized for MRI and CT generation~\cite{fetty2020latent}, MRI reconstruction for harmonization purposes~\cite{liu2023style}, and CT reconstruction~\cite{woodland2022evaluating}, consistently producing anatomically realistic and high-resolution images. More recently, VQGANs~\cite{esser2021taming} have been explored in medical imaging, combining vector quantization with adversarial training to better capture fine-grained structural details. VQGAN has been applied to brain tumor MRI reconstruction and has achieved superior performance compared to other advanced GAN variants and diffusion models, demonstrating its potential for complex medical image synthesis tasks~\cite{zhou2025generating}.

\subsection{GANs For PET Synthesis}
Several studies have demonstrated the feasibility of PET image synthesis using GAN-based methods. Since the most commonly used PET imaging technique relies on the radioactive tracer \textsuperscript{18}F-fluorodeoxyglucose (18F-FDG)~\cite{ul2012pet}, FDG PET synthesis has been the most extensively investigated. Wang \emph{et al.} proposed a progressive refinement cGAN framework to generate full-dose FDG PET from low-dose PET images, demonstrating improved quantitative accuracy and visual fidelity compared to conventional cGAN models~\cite{wang20183dd}. In another study, Wang \emph{et al.} developed a cGAN with multimodal inputs, achieving superior performance in synthesizing high-quality FDG PET images from low-dose FDG PET with MRI images compared to traditional multimodal GANs~\cite{wang20183d}. Kim \emph{et al.} compared multiple cGAN and WGAN architectures with varying objective functions and reported that the least-squares GAN (LSGAN)~\cite{mao2017least} yielded the best performance in estimating standardized uptake values (SUVs) from synthetic FDG PET images~\cite{kim2023deep}. Beyond paired data settings, Zotova \emph{et al.} applied CycleGAN to synthesize FDG PET from T1-weighted MRI and validated its clinical utility by confirming its effectiveness in detecting subtle epilepsy lesions~\cite{zotova2021gan}. 

Amyloid PET is another important imaging technique widely used to detect abnormal amyloid protein deposits associated with AD~\cite{chetelat2020amyloid,nordberg2004pet,altomare2021diagnostic}. With the growing availability of amyloid PET data, researchers have been exploring amyloid PET synthesis using GAN-based frameworks. Vega \emph{et al.} employed a cGAN to generate synthetic amyloid PET from structural T1-weighted MRI, producing results that closely matched ground-truth images across cognitive conditions~\cite{vega2024image}. Lan \emph{et al.} developed a multimodal cGAN framework that synthesized amyloid PET images from T1-weighted and FLAIR MRI, achieving superior quality~\cite{lan2021three}. In the unpaired domain, Pan \emph{et al.} established that CycleGAN could generate synthetic amyloid PET images with quality comparable to paired methods, underscoring its feasibility when aligned MRI–PET datasets are unavailable~\cite{pan2018synthesizing}. Tau PET is the latest and crucial PET technique in AD. However, research on tau PET synthesis remains scarce, largely due to the limited availability of tau PET datasets. Expanding GAN-based synthesis approaches to tau PET has a significant potential to advance AD research and clinical care, enabling cost-effective, non-invasive, and scalable access to tau pathology information.

\section{Methodology}
This study employs a 2D convolutional neural network (CNN)-based framework, MCR-VQGAN, to synthesize tau PET images from structural T1-weighted MRI.  As illustrated in Fig.~\ref{fig:architecture}~(a), the proposed architecture consists of two 2D CNNs: a generator network that performs MRI-to-PET translation and a discriminator network that enforces realistic image generation through adversarial training. A slice-by-slice 2D strategy is adopted instead of a full 3D implementation to enhance computational efficiency, critical for practical deployment in clinical settings. To evaluate the effectiveness of MCR-VQGAN, we compare its performance against established generative models, including cGAN~\cite{isola2017image}, WGAN with gradient penalty (WGAN-GP)~\cite{gulrajani2017improved}, CycleGAN~\cite{zhu2017unpaired}, and standard VQGAN~\cite{esser2021taming}. Furthermore, to validate the clinical utility of the synthetic tau PET images, we perform downstream AD classification tasks using a 2D CNN-based classifier. 

\begin{table}[!t]
\centering
\caption{Evaluation of vector quantization codebook parameters across codebook size ($K$) and embedding dimensionality ($D$). All configurations were trained with the full 500-epoch schedule.}
\label{tab:codebook_params}
\begin{tabular}{c c c c c}
\hline
$K$ & $D$ & MSE $\downarrow$ & PSNR (dB) $\uparrow$ & SSIM $\uparrow$ \\
\hline
256  & 256 & 0.0059 $\pm$ 0.0060 & 30.26 $\pm$ 4.29 & 0.9228 $\pm$ 0.0442 \\
256  & 512 & 0.0060 $\pm$ 0.0064 & 30.37 $\pm$ 4.45 & 0.9239 $\pm$ 0.0441 \\
512  & 256 & 0.0061 $\pm$ 0.0062 & 30.24 $\pm$ 4.40 & 0.9229 $\pm$ 0.0447 \\
512  & 512 & 0.0061 $\pm$ 0.0065 & 30.30 $\pm$ 4.42 & 0.9231 $\pm$ 0.0452 \\
1024 & 256 & 0.0061 $\pm$ 0.0067 & 30.30 $\pm$ 4.38 & 0.9235 $\pm$ 0.0447 \\
1024 & 512 & 0.0061 $\pm$ 0.0069 & 30.31 $\pm$ 4.41 & 0.9234 $\pm$ 0.0449 \\
\hline
\end{tabular}
\end{table}

\subsection{Datasets}
We obtained 222 pairs of structural T1-weighted MRI and tau PET scans from the Alzheimer’s Disease Neuroimaging Initiative 3 (ADNI-3) database (\underline{adni.loni.usc.edu}). The dataset included subjects across different clinical stages of AD:
\begin{itemize}
    \item Cognitive Normal (CN): 72 subjects
    \item Early Mild Cognitive Impairment (EMCI): 45 subjects
    \item Late Mild Cognitive Impairment (LMCI): 31 subjects
    \item Significant Memory Concern (SMC): 62 subjects
    \item Alzheimer’s Disease (AD): 12 subjects
\end{itemize}
All structural T1-weighted MRI scans were acquired on 3T scanners (GE, Philips, and Siemens) with parameters: repetition time (TR) = 2,300 ms, echo time (TE) = minimum full echo, inversion time (TI) = 900 ms, field of view = 208$\times$240$\times$256 mm, reconstruction resolution = 1$\times$1$\times$1 mm, and total acquisition time $\approx$ 6.20 min. All tau PET images were acquired using \textsuperscript{18}F-AV-1451 (flortaucipir) with parameters: injected dose = 370 MBq (10 mCi), minimum dose = 185 MBq (5 mCi), and post-injection imaging time = 75–105 min with 6$\times$5 min frames.

Each MRI–PET pair was obtained from the same individual, with MRI scans acquired on the same day or within $\le$100 days of the corresponding PET scan to ensure anatomical consistency. 

\subsection{Data Preprocessing}
We implemented separate preprocessing pipelines for tau PET and MRI scans to ensure accurate spatial alignment and quantification. Pipelines utilized multiple established neuroimaging tools, including FSL (FMRIB Software Library)~\cite{jenkinson2012fsl}, ANTs (Advanced Normalization Tools)~\cite{avants2009advanced}, and SPM (Statistical Parametric Mapping)~\cite{penny2011statistical}. 

For structural T1-weighted MRI, preprocessing steps included: (1) performing N4 bias field correction using ANTs for intensity uniformity, (2) aligning T1-weighted MRI to a structural template using FSL FLIRT, (3) performing intra-timepoint scaling to reduce variability across sessions, (4) timepoint transformation for longitudinal alignment using SPM to maintain temporal consistency, (5) performing nonlinear registration for precise spatial normalization via nearest-neighbor interpolation using ANTs, (6) applying a binary brain mask to the transformed images using FSL. 

For tau PET, preprocessing steps included: (1) performing motion correction using FSL MCFLIRT to align all PET images to the first timepoint and reduce movement artifacts, (2) temporal averaging of motion-corrected PET images to enhance signal quality, (3) affine registration of averaged PET images to corresponding T1-weighted MRI space using ANTs, (4) calculating Standardized Uptake Value Ratios (SUVRs) using cerebellar gray matter as a reference region.

These preprocessing steps ensured precise bidirectional mappings between the two imaging modalities for training and evaluation. All imaging data initially had a resolution of 256$\times$256$\times$170, with the axial view having a shape of 256$\times$170. For consistency, the axial slices were zero-padded to 256$\times$256, yielding the final volume size of 256$\times$256$\times$256.

\begin{table*}[!t]
\caption{Grid search results for generator loss weights ($\lambda_\text{adv} = 1$ fixed). Each configuration was trained for 100 epochs and evaluated on the test set.}
\label{tab:loss_weight_search}
\centering
\sisetup{separate-uncertainty=true, detect-weight=true}
\begin{tabular*}{\textwidth}{@{\extracolsep{\fill}}c c c 
    S[table-format=1.4(4)] 
    S[table-format=2.2(2)] 
    S[table-format=1.4(4)] @{}}
\toprule
{$\lambda_\text{rec}$} & {$\lambda_\text{perc}$} & {$\lambda_\text{VQ}$} & {MSE $\downarrow$} & {PSNR (dB) $\uparrow$} & {SSIM $\uparrow$} \\
\midrule
5  & 5  & 5  & 0.0062 \pm 0.0068 & 30.21 \pm 4.40 & 0.9218 \pm 0.0454 \\
5  & 5  & 10 & 0.0061 \pm 0.0066 & 30.32 \pm 4.48 & 0.9226 \pm 0.0450 \\
5  & 10 & 5  & 0.0060 \pm 0.0070 & 30.46 \pm 4.49 & 0.9248 \pm 0.0443 \\
5  & 10 & 10 & 0.0061 \pm 0.0066 & 30.30 \pm 4.41 & 0.9228 \pm 0.0442 \\
5  & 20 & 5  & 0.0059 \pm 0.0064 & 30.44 \pm 4.47 & 0.9255 \pm 0.0439 \\
5  & 20 & 10 & 0.0061 \pm 0.0068 & 30.35 \pm 4.44 & 0.9244 \pm 0.0441 \\
10 & 5  & 5  & 0.0061 \pm 0.0063 & 30.23 \pm 4.37 & 0.9231 \pm 0.0446 \\
10 & 5  & 10 & 0.0066 \pm 0.0071 & 29.96 \pm 4.48 & 0.9207 \pm 0.0461 \\
10 & 10 & 5  & 0.0066 \pm 0.0075 & 30.15 \pm 4.71 & 0.9210 \pm 0.0469 \\
10 & 10 & 10 & 0.0063 \pm 0.0069 & 30.12 \pm 4.40 & 0.9216 \pm 0.0448 \\
10 & 20 & 5  & 0.0062 \pm 0.0070 & 30.28 \pm 4.48 & 0.9228 \pm 0.0452 \\
10 & 20 & 10 & 0.0060 \pm 0.0064 & 30.33 \pm 4.38 & 0.9236 \pm 0.0441 \\
20 & 5  & 5  & 0.0061 \pm 0.0066 & 30.28 \pm 4.49 & 0.9225 \pm 0.0451 \\
20 & 5  & 10 & 0.0061 \pm 0.0064 & 30.13 \pm 4.22 & 0.9223 \pm 0.0438 \\
20 & 10 & 5  & 0.0062 \pm 0.0067 & 30.24 \pm 4.43 & 0.9231 \pm 0.0446 \\
20 & 10 & 10 & 0.0062 \pm 0.0068 & 30.16 \pm 4.36 & 0.9229 \pm 0.0449 \\
20 & 20 & 5  & 0.0061 \pm 0.0067 & 30.34 \pm 4.49 & 0.9236 \pm 0.0444 \\
20 & 20 & 10 & 0.0062 \pm 0.0069 & 30.26 \pm 4.34 & 0.9229 \pm 0.0446 \\
\bottomrule
\end{tabular*}
\end{table*}

\subsection{MCR-VQGAN Key Components}
MCR-VQGAN enhances the original VQGAN architecture by incorporating multi-scale convolutions, ResNet blocks~\cite{he2016deep}, and CBAM modules~\cite{woo2018cbam}. These components work synergistically to improve the model's ability to learn complex nonlinear mappings from structural T1-weighted MRI to tau PET. Together, these components enable the model to capture better anatomical and pathological details while also preserving the global anatomical context essential for AD imaging.

\subsubsection{Multi-Scale Convolution}
Conventional VQGANs typically employ uniform 3$\times$3 kernels, constraining the receptive field and limiting the capture of long-range spatial patterns. In neuroimaging, this is problematic because both local detail and global structure are clinically important. To address this limitation, as shown in Fig.~\ref{fig:architecture}~(b), MCR-VQGAN introduces parallel convolutional branches with kernels of sizes 3$\times$3, 5$\times$5, and 7$\times$7, each designed to capture features at different spatial scales:
\begin{itemize}
    \item 3$\times$3 Kernels: Small details and local textures
    \item 5$\times$5 Kernels: Medium-scale features for structural and anatomical patterns
    \item 7$\times$7 Kernels: Broader structural context and global spatial relationships
\end{itemize}
Outputs from these parallel branches are concatenated and processed through a 1$\times$1 convolution for adaptive feature fusion. The multi-scale convolutions are implemented within the encoder and ResNet blocks before vector quantization, enriching the latent representation. In contrast, as shown in Fig.~\ref{fig:architecture}~(c), the decoder and post-quantization ResNet blocks use standard 3$\times$3 kernels, prioritizing efficient reconstruction.

\subsubsection{ResNet Block}
Training deep generative models often leads to vanishing gradients and degradation of fine structural details. To overcome these challenges, MCR-VQGAN incorporates ResNet blocks~\cite{he2016deep}. In the encoder, ResNet blocks are placed after the last encoding stage. Each ResNet block consists of a multi-scale convolutional layer, instance normalization, ReLU, dropout ($p=0.5$), a second multi-scale convolutional layer, and final instance normalization. In the decoder, ResNet blocks are located before the upsampling stage, using standard 3$\times$3 convolutions instead of multi-scale convolutions to balance reconstruction quality with computational efficiency.

\subsubsection{Convolutional Block Attention Module (CBAM)}
To enhance feature learning, MCR-VQGAN incorporates CBAM~\cite{woo2018cbam} across the encoder, decoder, and ResNet blocks. As depicted in Fig.~\ref{fig:architecture}~(d), CBAM applies channel and spatial attention sequentially. The channel attention module processes feature maps via parallel global average pooling and global max pooling, followed by a shared multi-layer perceptron (MLP) to generate channel-wise attention weights. The spatial attention module then computes spatial attention weights by applying channel-wise average and max pooling, concatenating the results, and processing through a 7$\times$7 convolutional layer. This mechanism enables the network to focus on clinically relevant regions and features critical for tau pathology.

\subsection{Generator Architecture}
As presented in Fig.~\ref{fig:architecture}~(a), the generator consists of three primary components: an encoder for feature extraction, a vector quantization codebook for discrete representation learning, and a decoder for high-fidelity image reconstruction.

\subsubsection{Encoder}
The encoder begins with an initial 7$\times$7 convolutional layer for initial feature extraction, followed by progressive downsampling stages with channel dimensions of 64, 128, 256, and 512, respectively. Each downsampling stage integrates multi-scale convolutional blocks, instance normalization, ReLU, and a CBAM module, capturing both local details and global anatomical context. After the downsampling path, the output features are processed by six encoding ResNet blocks to refine the latent representation. Each ResNet block consists of a multi-scale convolution, instance normalization, ReLU, dropout ($p=0.5$), a second multi-scale convolution, and a final instance normalization. This design improves gradient flow, mitigates over-smoothing, and enhances the preservation of clinically relevant features.

\begin{table*}[!t]
\caption{Ablation study of the proposed MCR-VQGAN on tau PET synthesis, evaluating the incremental impact of multi-scale convolutions (MC), ResNet blocks (RB), and Convolutional Block Attention Modules (CBAM).}
\label{tab:ablation}
\centering
\sisetup{separate-uncertainty=true, detect-weight=true} 
\begin{tabular*}{\textwidth}{@{\extracolsep{\fill}}l 
    S[table-format=1.4(4)] 
    S[table-format=2.2(2)] 
    S[table-format=1.4(4)] @{}}
\toprule
Method & {MSE $\downarrow$} & {PSNR (dB) $\uparrow$} & {SSIM $\uparrow$} \\
\midrule
Standard VQGAN                     & 0.0069 \pm 0.0065 & 29.22 \pm 3.71 & 0.9124 \pm 0.0479 \\
VQGAN + Multi-Conv                 & 0.0061 \pm 0.0066 & 29.99 \pm 3.96 & 0.9198 \pm 0.0475 \\
VQGAN + Multi-Conv + ResNet Blocks & 0.0059 \pm 0.0060 & 30.29 \pm 4.38 & 0.9234 \pm 0.0460 \\
\textbf{MCR-VQGAN (Full)} &
{\bfseries \num{0.0056 \pm 0.0061}} &
{\bfseries \num{30.65 \pm 4.47}} &
{\bfseries \num{0.9263 \pm 0.0469}} \\
\bottomrule
\end{tabular*}
\end{table*}

\begin{figure*}[!t]
\centering
\includegraphics[width=0.9\textwidth]{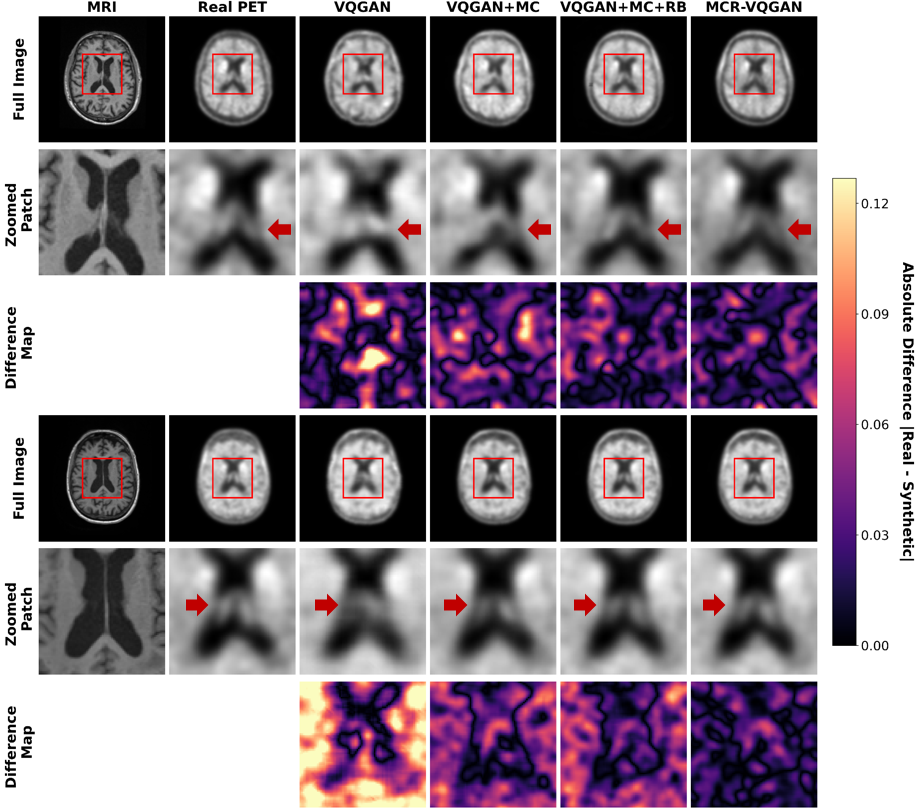}
\caption{Qualitative comparison of the ablation study on MCR-VQGAN components for T1-weighted MRI to tau PET synthesis. Representative axial slices illustrate the progressive performance enhancement from the baseline VQGAN to the full MCR-VQGAN (incorporating MC, RB, and CBAM). Each row presents the synthesized PET images, magnified views of key anatomical regions, and corresponding absolute difference maps compared to the ground-truth PET.}
\label{fig:ablation_comp}
\end{figure*}

\subsubsection{Vector Quantization Codebook}
The vector quantization component bridges the encoder and decoder. The encoded features are processed through a 1$\times$1 convolutional layer to map the encoded features to the quantization embedding space. Afterwards, the processed features pass through the vector quantization codebook, which contains $K=1024$ embedding vectors of dimension $D=512$, and are updated with their closest matching codebook vector. The selected embeddings are then mapped back into the latent space through another 1$\times$1 convolutional layer. The codebook is optimized using an exponential moving average (EMA) update with a decay factor of $\beta=0.99$, providing stable and consistent codebook learning. This quantization step enables the model to represent structural information in a discrete space, thereby enhancing the fidelity and robustness of cross-modality synthesis.

\subsubsection{Decoder}
The decoder reconstructs high-fidelity tau PET images from the quantized latent features. Its architecture is symmetric to the encoder, beginning with six decoding ResNet blocks to refine the latent representation. Unlike the encoder, each block uses standard 3$\times$3 convolutions for efficient reconstruction, combined with instance normalization, ReLU, dropout ($p=0.5$), and a second 3$\times$3 convolution. Following the ResNet blocks, three progressive upsampling stages restore spatial resolution to the original image size, with channel dimensions decreasing from 512 to 256, 128, and 64. Each stage employs a transposed convolution, followed by a standard 3$\times$3 convolutional block, instance normalization, ReLU, and a CBAM module to enhance reconstruction quality. Finally, a 7$\times$7 convolutional layer with Tanh activation function maps the output to a single-channel tau PET image, with intensity values scaled to the range [-1, 1]. This symmetric encoder–decoder design ensures a robust mapping from the latent space back to the image space while optimizing for reconstruction quality.

\subsection{Discriminator Architecture}
A PatchGAN discriminator~\cite{isola2017image} is adopted for adversarial training. The network consists of three convolutional layers with channel dimensions of 64, 128, and 256. Each layer uses 4$\times$4 kernels with a stride of 2, followed by instance normalization and LeakyReLU. The final layer outputs a single channel prediction map that classifies overlapping image patches as real or synthetic. This patch-based approach encourages the generator to produce realistic local details while maintaining global consistency.

\subsection{AD Classifier Architecture}
To evaluate the clinical utility of the synthesized tau PET images beyond pixel-level similarity, a downstream AD classification task was designed. While conventional metrics like MSE, PSNR, and SSIM quantify reconstruction accuracy, they do not necessarily confirm the preservation of subtle pathological biomarkers. To address this, we designed an experiment to test whether our synthetic images could serve as a viable surrogate for real images in a clinical task. The classifier was trained exclusively on real tau PET images from the training set and then evaluated on real and synthetic tau PET images separately. This approach allows for a direct assessment of whether the synthetic data retain the critical diagnostic information required for the task.

A custom 2D CNN-based classifier was developed for downstream AD classification tasks. The model consists of five convolutional blocks with progressively increasing channel dimensions of 16, 32, 64, 128, and 256, followed by a fully connected classifier. Each convolutional block contains two 3$\times$3 convolutional layers followed by batch normalization and ReLU. Max-pooling is applied at the end of each of the first four convolutional blocks to reduce spatial resolution. A global average pooling layer aggregates spatial features into a compact vector, which is passed through a fully connected classifier with two hidden layers (128 and 64 units) with dropout regularization ($p=0.3$). The output layer produces a binary probability for distinguishing CN versus MCI/AD, with a classification threshold of 0.5.

\subsection{Objective Functions}
The objective functions used to train the proposed MCR-VQGAN include two composite loss functions: one for the generator ($G$) and another for the discriminator ($D$). 

\subsubsection{Generator Loss}
The generator loss is a weighted sum of four distinct losses: reconstruction, perceptual, adversarial, and vector quantization. Given the structural T1-weighted MRI $x$, the real tau PET image $y$, and the synthetic tau PET image $\hat{y}$, the adversarial loss encourages the generator to produce realistic images:
\begin{equation}\label{eq_1}
\begin{split}
    \mathcal{L}_{adv}(G,D)=-\mathbb{E}_{x,\hat{y}}\left[\log D(x,\hat{y})\right] 
\end{split}
\end{equation}
The reconstruction loss $\mathcal{L}_{rec}$ enforces pixel-wise similarity between real and generated tau PET images:
\begin{equation}\label{eq_2}
    \mathcal{L}_{rec}(G)=\mathbb{E}_{x,y}\left[\|y-\hat{y}\|_1 \right]
\end{equation}
The perceptual loss $\mathcal{L}_{perc}$ uses a pretrained VGG16 network to preserve high-level structural and textural characteristics:
\begin{equation}\label{eq_3}
    \mathcal{L}_{perc}(G)=\sum_j\frac{1}{H_jW_j}\sum_{h,w}\|\phi_j(y)_{h,w}-\phi_j(\hat{y})_{h,w}\|_2^2
\end{equation}
, where $H_j$ and $W_j$ represent the spatial dimensions of the feature map at layer $j$, and $\phi(\cdot)$ represents feature maps from the $j$th layer of pretrained VGG16. The vector quantization loss $\mathcal{L}_{VQ}$ encourages encoder outputs to align with the nearest codebook embeddings:
\begin{equation}\label{eq_4}
    \mathcal{L}_{VQ}(G)=\|sg[E(x)]-z_q\|_2^2+\beta\|sg[z_q]-E(x)\|_2^2
\end{equation}
, where $E(x)$ represents the encoder output, $z_q$ represents the selected codebook vector, $sg[\cdot]$ denotes the stop-gradient operator, and $\beta$ represents the commitment weight. The complete generator loss is:
\begin{equation}\label{eq_5}
    \mathcal{L}_{G}=\lambda_{adv}\mathcal{L}_{adv}+\lambda_{rec}\mathcal{L}_{rec}+\lambda_{perc}\mathcal{L}_{perc}+\lambda_{VQ}\mathcal{L}_{VQ}
\end{equation}
, where $\lambda_{adv}$, $\lambda_{rec}$, $\lambda_{perc}$, and $\lambda_{VQ}$ are the weighting coefficients for each loss component.

\subsubsection{Discriminator Loss}
The discriminator maximizes its ability to distinguish between real and synthetic tau PET images by minimizing:
\begin{equation}\label{eq_6}
    \mathcal{L}_{D}=-\mathbb{E}_{x,y}\left[\log D(x,y)\right]-\mathbb{E}_{x,\hat{y}}\left[\log (1-D(x,\hat{y}))\right] 
\end{equation}
To further stabilize training, we apply R1 gradient penalty regularization~\cite{mescheder2018training}:
\begin{equation}\label{eq_r1}
    R_1 = \frac{\gamma}{2} \mathbb{E}_{x,y}\left[\|\nabla_{y} D(x,y)\|^2\right]
\end{equation}
, where $\nabla_y D(x,y)$ denotes the gradient of the discriminator output with respect to the real tau PET image $y$, and $\gamma=10$ is the penalty coefficient. The final discriminator loss becomes:
\begin{equation}\label{eq_d_final}
    \mathcal{L}_{D,total} = \mathcal{L}_{D} + R_1
\end{equation}

\subsubsection{AD Classifier Loss}
For downstream AD classification tasks, focal loss \cite{lin2017focal} is utilized to address class imbalance between the diagnostic labels (CN vs. MCI/AD). The focal loss is defined as:
\begin{equation}\label{eq_7}
    \mathcal{L}_{focal} = -\alpha(1-p_t)^{\gamma}\log (p_t)
\end{equation}
, where $p_t$ is the predicted probability for the true class, $\alpha$ is the weighting factor for class balance, and $\gamma$ is the focusing parameter.

\subsection{Training Pipeline}
The overall training pipeline consisted of two main phases: (1) MCR-VQGAN and baseline model training, and (2) downstream AD classifier training. This study used paired structural T1-weighted MRI and tau PET volumes from the ADNI-3 dataset. The dataset was randomly split into training ($n = 178$) and test ($n = 44$) sets, maintaining proportional representation of clinical diagnoses (Table~\ref{tab:dataset_distribution}). From each paired volume, 14 axial slices were extracted from the central brain region. Slice selection targeted the central 100 axial slices (indices 76–176), selecting every 10th slice plus four extra slices (124, 128, 130, 132) to capture key anatomy. Experiments were run on a workstation with a single NVIDIA L4 GPU (24 GB memory).

\begin{table}[!ht]
\caption{Distribution of Clinical Diagnoses in the Training and Test Sets}
\label{tab:dataset_distribution}
\centering
\begin{tabular*}{\columnwidth}{l @{\extracolsep{\fill}} ccc}
\toprule
\textbf{Diagnosis} & \textbf{Training Set} & \textbf{Test Set} & \textbf{Total} \\
\midrule
Cognitive Normal (CN)       & 58 (32.6\%) & 14 (31.8\%) & 72 (32.4\%) \\
Early MCI (EMCI)            & 36 (20.2\%) & 9 (20.5\%)  & 45 (20.3\%) \\
Late MCI (LMCI)             & 25 (14.0\%) & 6 (13.6\%)  & 31 (14.0\%) \\
Sig. Memory Concern (SMC)   & 50 (28.1\%) & 12 (27.3\%) & 62 (27.9\%) \\
Alzheimer's Disease (AD)    & 9 (5.1\%)   & 3 (6.8\%)   & 12 (5.4\%)  \\
\midrule
\textbf{Total Subjects}     & \textbf{178 (100\%)} & \textbf{44 (100\%)} & \textbf{222 (100\%)} \\
\bottomrule
\end{tabular*}
\end{table}

\subsubsection{Generative Model Training}
MCR-VQGAN and baseline models, including cGAN, CycleGAN, WGAN-GP, and standard VQGAN, were trained for 500 epochs with an initial learning rate of 0.0002, which was decayed using a cosine annealing scheduler~\cite{loshchilov2016sgdr}. To address training instability and prevent discriminator overfitting, common challenges in GAN training, we applied several stabilization techniques. First, one-sided label smoothing was used during discriminator training, replacing hard labels of 1 for real images with values sampled uniformly from $[0.9, 1.0]$. Second, for all VQGAN-based models, Gaussian noise was added to generated images with variance $\sigma^2=0.1\times(1-epoch/500)$, gradually decreasing from 0.1 to 0 to stabilize early training while capturing original details in later epochs.

Hyperparameter tuning is a critical step in ensuring optimal performance in deep learning models. Since one of the primary tasks of this study is to evaluate the fidelity of tau PET synthesis among various GAN architectures, shared training hyperparameters, including learning rate (0.0002), number of training epochs (500), and optimizer momentum parameters ($\beta_1 = 0.5$, $\beta_2 = 0.999$), were kept constant across all models to ensure fair comparisons. Batch sizes were determined by the maximum GPU memory capacity available for each architecture: 2 for MCR-VQGAN, CycleGAN, and standard VQGAN, and 4 for WGAN-GP and cGAN. Baseline models (cGAN, CycleGAN, and WGAN-GP) were trained using default configurations as described in their original publications. For the proposed MCR-VQGAN, model-specific hyperparameters, including codebook size ($K$), embedding dimensionality ($D$), and generator loss weights, were optimized through systematic evaluations.

The parameters defining the vector quantization codebook were evaluated across $K \in \{256, 512, 1024\}$ and $D \in \{256, 512\}$, yielding six configurations. The generator loss weights (Eq.~\eqref{eq_5}) were optimized via grid search over $\lambda_\text{rec} \in \{5, 10, 20\}$, $\lambda_\text{perc} \in \{5, 10, 20\}$, and $\lambda_\text{VQ} \in \{5, 10\}$, with $\lambda_\text{adv} = 1$ fixed, yielding 18 configurations. Each configuration was trained for 100 epochs and evaluated using MSE, PSNR, and SSIM on the test set. The selected configuration was trained for the full 500 epochs.

\begin{table*}[!t]
\caption{Quantitative comparison of MCR-VQGAN against cGAN, WGAN-GP, CycleGAN, and VQGAN for tau PET synthesis.}
\label{tab:comparison}
\centering
\sisetup{separate-uncertainty=true, detect-weight=true}
\begin{tabular*}{\textwidth}{@{\extracolsep{\fill}}l 
    S[table-format=1.4(4)] 
    S[table-format=2.2(2)] 
    S[table-format=1.4(4)] @{}}
\toprule
Method & {MSE $\downarrow$} & {PSNR (dB) $\uparrow$} & {SSIM $\uparrow$} \\
\midrule
cGAN      & 0.0071 \pm 0.0079 & 29.60 \pm 4.33 & 0.9119 \pm 0.0502 \\
WGAN-GP   & 0.0067 \pm 0.0075 & 29.94 \pm 4.36 & 0.9179 \pm 0.0488 \\
CycleGAN  & 0.0112 \pm 0.0098 & 26.78 \pm 3.26 & 0.8830 \pm 0.0525 \\
VQGAN     & 0.0069 \pm 0.0065 & 29.22 \pm 3.71 & 0.9124 \pm 0.0479 \\
\textbf{MCR-VQGAN (Full)} &
{\bfseries \num{0.0056 \pm 0.0061}} &
{\bfseries \num{30.65 \pm 4.47}} &
{\bfseries \num{0.9263 \pm 0.0469}} \\
\bottomrule
\end{tabular*}
\end{table*}

\begin{figure*}[!t]
\centering
\includegraphics[width=0.9\textwidth]{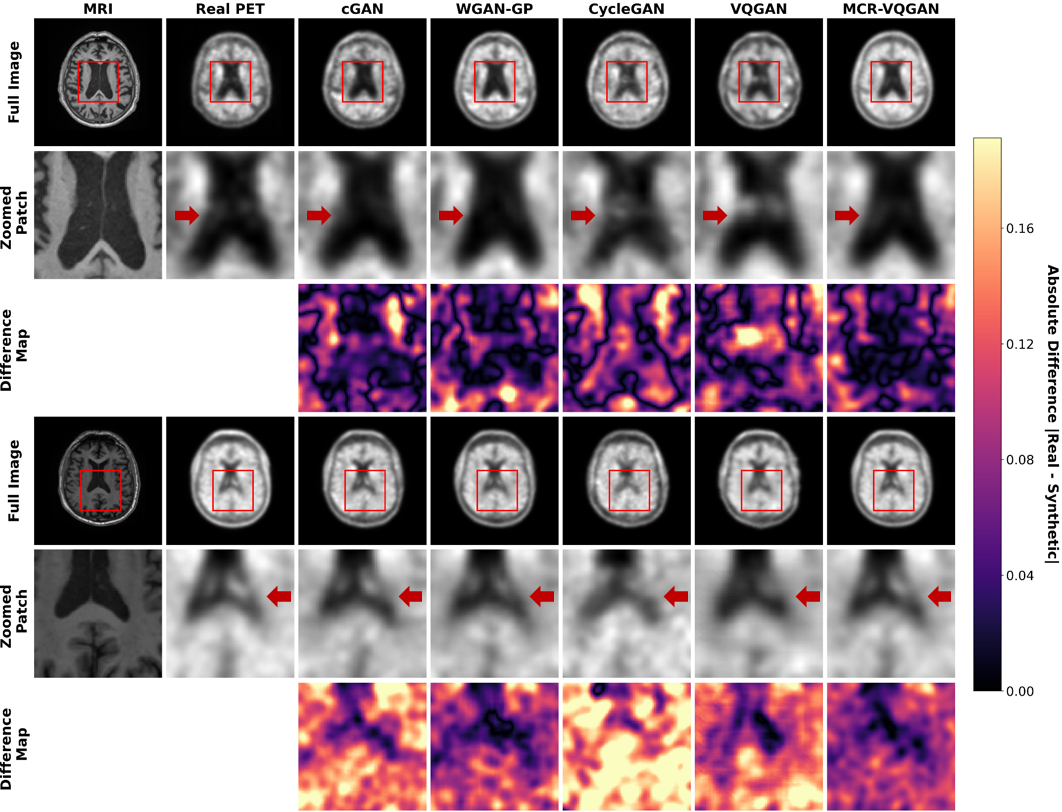}
\caption{Qualitative comparison of five generative models for T1-weighted MRI to tau PET synthesis. Representative axial slices demonstrate that MCR-VQGAN generates high-fidelity PET images with the highest structural consistency and minimal artifacts. For each slice, the panels illustrate the full synthesized volume, a magnified view of the indicated ROI (red box), and the corresponding voxel-wise absolute difference map compared to the ground-truth PET.}
\label{fig:baseline_comp}
\end{figure*}

\subsubsection{AD Classifier Training}
To further evaluate the clinical fidelity of MCR-VQGAN, downstream AD classification tasks were performed to determine whether the synthesized tau PET images preserved diagnostically relevant biomarkers. The classifier was trained and evaluated using the same training and test sets. For preprocessing, all real tau PET volumes were cropped to the middle 100 axial slices (indices 76–176) to ensure consistency with the input format of MCR-VQGAN. For test subjects, the corresponding 100 MRI slices were translated into synthetic tau PET images using the trained MCR-VQGAN model. For this binary task, SMC and CN subjects were grouped as the CN class, while EMCI, LMCI, and AD subjects were grouped as the MCI/AD class, yielding 26 CN and 18 MCI/AD subjects in the test set. Subject-level predictions were obtained by averaging the classifier output probabilities across all 100 axial slices (indices 76–176) for each subject and applying a decision threshold of 0.5.

To mitigate label imbalance (CN vs. MCI/AD), every other slice was sampled for CN subjects, while all slices were included for MCI/AD cases. A custom 2D CNN classifier was designed with binary output. The network was trained for 200 epochs with an initial learning rate of 0.0002, decayed using an exponential scheduler (gamma = 0.98), and a batch size of 8. Optimization was performed using the Adam optimizer. To improve the generalizability of the model, data augmentations, including random rotations, flips, and affine transformations, were applied during training. To address residual class imbalance and emphasize difficult examples, the classifier was trained with focal loss ($\alpha=0.25$, $\gamma=2.0$), which adaptively downweights easy negatives and forces the model to focus on underrepresented or challenging samples.

\subsection{Evaluation Metrics}
To quantitatively evaluate the proposed MCR-VQGAN, we employed three standard image quality metrics: Mean Squared Error (MSE), Peak Signal-to-Noise Ratio (PSNR), and Structural Similarity Index Measure (SSIM). MSE measures the pixel-wise intensity difference, while PSNR assesses the reconstruction quality relative to the peak signal. SSIM evaluates the perceptual structural similarity between the synthetic and ground-truth images. All image quality metrics (PSNR and SSIM) were computed using a standardized data range of 2.0, corresponding to the normalized image intensity range of $[-1, 1]$. This fixed data range ensures consistent and comparable evaluation across all subjects and models.

For the downstream classification task, we used accuracy to evaluate the model's ability to distinguish between CN and MCI/AD subjects. Accuracy is calculated as the ratio of correct predictions to the total number of predictions.

\subsection{Braak Staging and Regional SUVR Analysis}
To evaluate the clinical relevance of the synthesized tau PET images beyond pixel-level similarity metrics, we performed a regional quantitative analysis based on the Braak staging framework. Braak staging describes how tau pathology spreads through the brain---from the entorhinal cortex and hippocampus (stages I/II), through the inferior temporal and limbic cortices (stages III/IV), to wider association and primary sensory cortices (stages V/VI)~\cite{braak1991neuropathological, braak2006staging}. Because this hierarchical spatial progression underlies clinical PET-based tau assessment~\cite{scholl2016pet, therriault2022biomarker}, agreement within Braak-defined regions provides a clinically meaningful measure of synthesis fidelity.

Since our tau PET volumes are normalized to the $[-1, 1]$ range for stable GAN training rather than scaled to absolute flortaucipir concentration, published SUVR thresholds for categorical Braak staging~\cite{scholl2016pet, therriault2022biomarker} are not directly applicable. We therefore performed the Braak-framework evaluation based on regional SUVR agreement, which preserves the same ROI definitions and reference-region normalization as PET-based staging while remaining compatible with our normalized image representation. ROIs were defined using FreeSurfer's Desikan-Killiany atlas (aparc+aseg) and grouped into three hierarchical composites following established protocols: Braak I/II (entorhinal cortex, hippocampus), Braak III/IV (parahippocampal, fusiform, inferior/middle temporal gyri, amygdala, adjacent limbic regions), and Braak V/VI (remaining cortical regions)~\cite{scholl2016pet, therriault2022biomarker}. An additional Other ROI served as a control, and bilateral cerebellar gray matter served as the SUVR reference.

\subsection{SUVR Computation and Statistical Analysis}
Real and synthetic tau PET volumes were linearly rescaled from $[-1, 1]$ to $[0, 1]$ before computing the SUVR-equivalent ratio as the mean ROI intensity divided by the mean cerebellar gray matter intensity. FreeSurfer parcellations were resampled to the synthesis volume dimensions $(256 \times 256 \times 101)$ via nearest-neighbor interpolation using FSL~\cite{jenkinson2012fsl}. Among the 44 test subjects, 37 had FreeSurfer parcellations available in ADNI; therefore, the remaining 7 were excluded. Agreement between real and synthetic SUVR was quantified using Pearson correlation, paired $t$-tests, and the intraclass correlation coefficient (ICC) under a two-way random-effects model for absolute agreement of single measurements, with ICC values classified~\cite{koo2016guideline} as poor $(< 0.50)$, moderate $(0.50\text{--}0.75)$, good $(0.75\text{--}0.90)$, or excellent $(> 0.90)$.

\section{Results and Analysis}
To quantitatively evaluate the proposed MCR-VQGAN framework, synthesized tau PET images were compared against their corresponding ground-truth scans using three standard metrics: MSE to assess pixel-wise reconstruction accuracy, PSNR to measure image fidelity, and SSIM to quantify perceptual similarity and the preservation of structural details. Beyond reconstruction quality, the clinical utility of the synthesized images was evaluated through a downstream binary classification task (CN vs. MCI/AD). This experiment assessed whether the MCR-VQGAN framework effectively preserves the subtle pathological biomarkers essential for diagnostic decision-making.

\subsection{Hyperparameter Optimization Results}
The evaluation of vector quantization codebook revealed that all six configurations achieved comparable performance as described in Table~\ref{tab:codebook_params}, indicating robustness to codebook size. We adopted $K = 1024$ and $D = 512$ following the original VQGAN framework~\cite{esser2021taming}, as the larger codebook provides greater representational capacity for capturing diverse tau deposition patterns. The embedding dimension $D = 512$ was chosen to match the encoder's output channel dimension, enabling direct mapping without dimensionality reduction.

For generator loss weights, 18 configurations were evaluated on the test set after 100 epochs (Table~\ref{tab:loss_weight_search}) as an initial screening step. Because differences between top-ranked configurations were small (SSIM range: $0.9207$--$0.9255$), the final configuration $\lambda_\text{rec} = 10$, $\lambda_\text{perc} = 10$, $\lambda_\text{VQ} = 5$ was selected based on a combination of validation metrics and balanced loss weighting, which tends to promote stable convergence over extended training. This configuration was subsequently retrained for the full 500-epoch schedule, during which no adversarial collapse was observed.

\subsection{Ablation Study Results}
The performance of MCR-VQGAN was evaluated through a stepwise ablation study to quantify the individual contributions of multi-scale convolutions (MC), ResNet blocks (RB), and CBAM. Quantitative results are summarized in Table~\ref{tab:ablation}, while Fig.~\ref{fig:ablation_comp} presents qualitative comparisons alongside their corresponding absolute difference maps. Consistent performance gains were observed across all metrics with the integration of each component. As shown in Table~\ref{tab:ablation}, the baseline VQGAN exhibited the lowest performance (MSE: 0.0069, PSNR: 29.22 dB, SSIM: 0.9124). Integration of MC and RB blocks provided steady improvements, while the full MCR-VQGAN achieved superior performance: the lowest MSE (0.0056) and the highest PSNR (30.65 dB) and SSIM (0.9263). These quantitative improvements are visually demonstrated in Fig.~\ref{fig:ablation_comp}, where the full model progressively sharpens the anatomical structures and minimizes residual errors in the difference maps. These findings confirm that the synergy between the proposed architectural enhancements is critical for high-fidelity tau PET synthesis.

\subsection{Comparison Study Results}
To evaluate its performance, MCR-VQGAN was compared against several established generative models: cGAN, WGAN-GP, CycleGAN, and the standard VQGAN. As summarized in Table~\ref{tab:comparison}, MCR-VQGAN consistently outperformed all baselines in every metric. In particular, MCR-VQGAN surpassed the strongest baseline, WGAN-GP (MSE: 0.0067, PSNR: 29.94 dB, SSIM: 0.9179), by achieving a 16.4\% reduction in MSE (0.0056) and superior structural fidelity (PSNR: 30.65 dB and SSIM: 0.9263). This quantitative superiority is visually substantiated in Fig.~\ref{fig:baseline_comp}. The magnified patches revealed that MCR-VQGAN produced sharper and more anatomically faithful details, while the absolute difference maps demonstrated significantly lower residual errors across the entire brain compared to all baseline architectures.

\begin{figure}[!t]
\centering
\includegraphics[width=\linewidth]{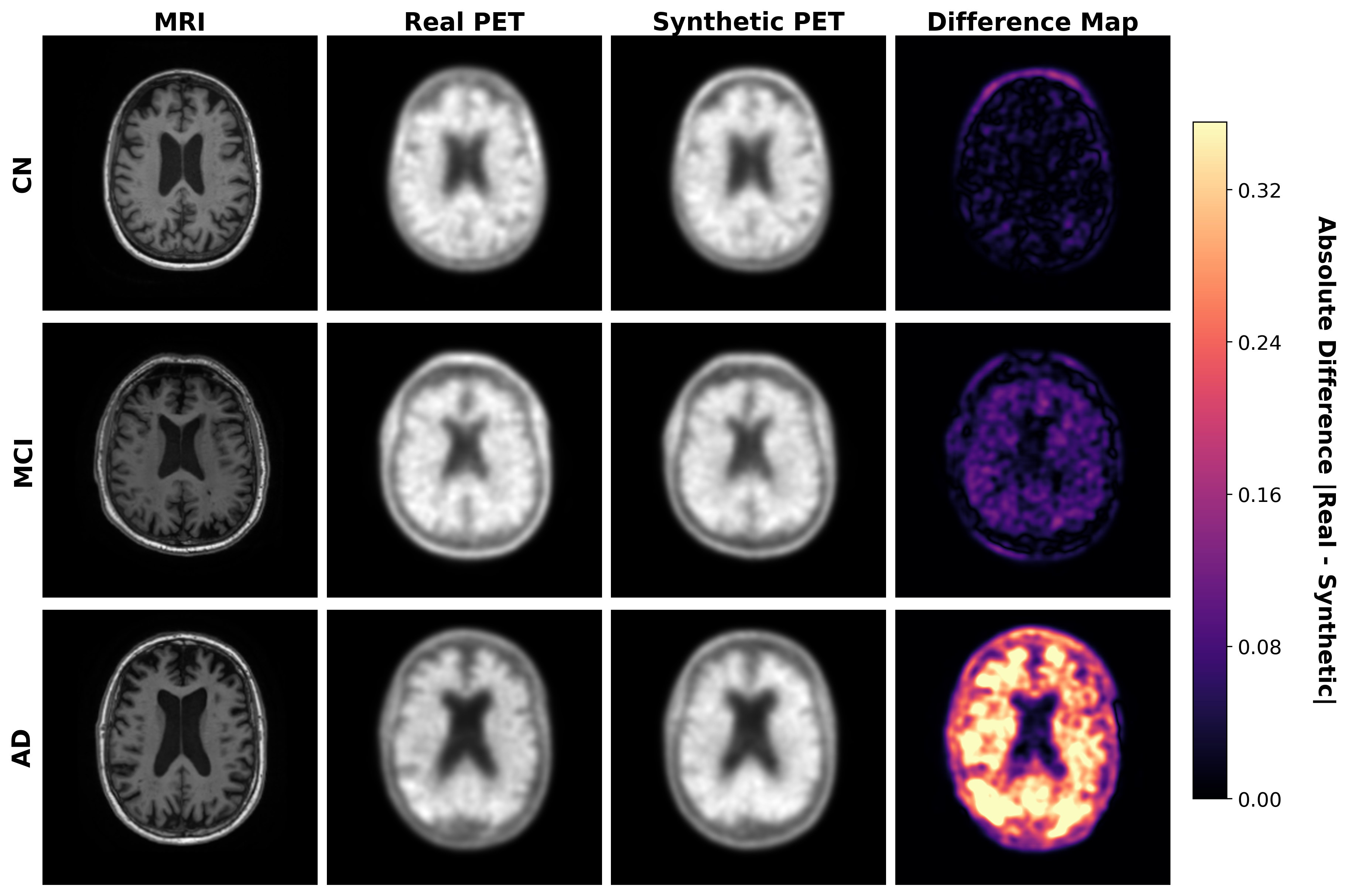}
\caption{Representative results of MRI-to-tau PET synthesis across different diagnostic groups (CN, MCI, and AD). The visual comparison highlights the model’s ability to reconstruct tau distribution patterns across the AD continuum. From left to right: T1-weighted MRI, ground-truth tau PET, MCR-VQGAN output, and voxel-wise absolute difference maps. Note the increased structural complexity and synthesis challenge as disease severity progresses from CN to AD.}
\label{fig:group_comp}
\end{figure}

Furthermore, synthesis performance was stratified across diagnostic groups (CN, MCI, and AD) to assess the model's robustness to pathology. As shown in Table~\ref{tab:cog_comparison}, synthesis fidelity was inversely correlated with disease severity: CN subjects achieved the highest fidelity, followed by MCI, while AD subjects exhibited the lowest performance. This trend, illustrated in Fig.~\ref{fig:group_comp}, suggests that the advanced structural and pathological alterations associated with AD present an increased challenge for cross-modality synthesis, necessitating further investigation into pathological feature preservation.

\begin{table}[!t]
\caption{Performance evaluation of MCR-VQGAN-based PET synthesis across different cognitive states. CN includes SMC subjects, and MCI includes EMCI and LMCI subjects. Values are reported as mean (SD).}
\label{tab:cog_comparison}
\centering
\begin{tabular*}{\columnwidth}{@{\extracolsep{\fill}} l c c c @{}}
\toprule
Clinical Diagnosis & MSE $\downarrow$ & PSNR (dB) $\uparrow$ & SSIM $\uparrow$ \\
\midrule
CN (n=26)  & 0.0042 (0.0043) & 31.53 (4.12) & 0.9356 (0.0343) \\
MCI (n=15) & 0.0059 (0.0061) & 30.32 (4.33) & 0.9212 (0.0516) \\
AD (n=3)   & 0.0163 (0.0080) & 24.74 (3.13) & 0.8717 (0.0693) \\
\bottomrule
\end{tabular*}
\end{table}

\subsection{Downstream AD Classification Task}
To evaluate the clinical utility of the synthesized images, a downstream classification experiment was conducted to distinguish CN from MCI/AD subjects. A classifier trained exclusively on real tau PET images was evaluated separately on a test set of real scans and their corresponding MCR-VQGAN-synthesized counterparts. As summarized in Table~\ref{table_3}, the classifier achieved comparable performance across both datasets, yielding an overall accuracy of 63.64\% (28/44) on real images and 65.91\% (29/44) on synthetic images. Notably, for the CN subgroup, the accuracy was identical for both modalities at 73.08\% (19/26). In the MCI/AD subgroup, the synthetic images achieved a slightly higher accuracy of 55.56\% (10/18) compared to 50.00\% (9/18) for the real images. These results suggest that MCR-VQGAN preserves diagnostically relevant features and pathological biomarkers supporting AD classification, indicating that the synthetic images retain key diagnostic information.

\begin{table}[!t]
\caption{Downstream classification performance (CN vs. MCI/AD) for a 2D CNN trained on real tau PET and evaluated on real and MCR-VQGAN–generated images.}
\label{table_3}
\centering
\begin{tabular*}{\columnwidth}{@{\extracolsep{\fill}} l c c @{}}
\toprule
\textbf{Category} & \textbf{Real tau PET} & \textbf{Synthetic tau PET} \\
\midrule
Overall Accuracy & 28 / 44 (63.64\%) & 29 / 44 (65.91\%) \\
CN Accuracy      & 19 / 26 (73.08\%) & 19 / 26 (73.08\%) \\
MCI/AD Accuracy  & 09 / 18 (50.00\%) & 10 / 18 (55.56\%) \\
\bottomrule
\end{tabular*}
\end{table}

\subsection{Regional SUVR Comparison Across Braak ROIs}
\begin{figure*}[!t]
\centering
\includegraphics[width=0.9\textwidth]{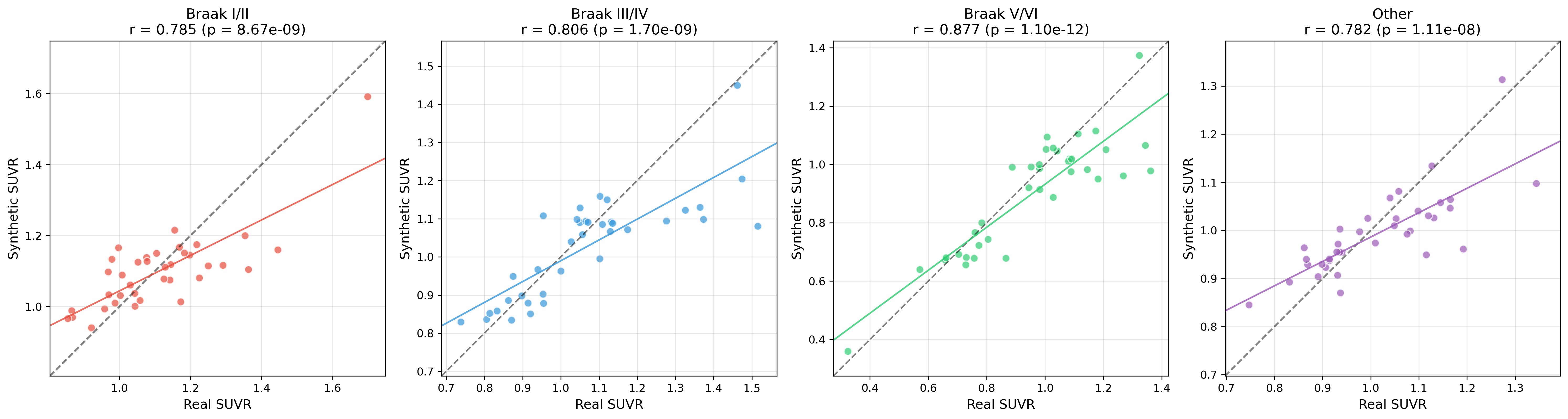}
\caption{Scatter plots of real versus MCR-VQGAN-synthesized SUVR values across Braak-defined ROIs ($n = 37$). Each point represents one subject. The dashed line indicates perfect agreement (identity line), and the solid line represents the linear regression fit. Pearson correlation coefficients are reported for each ROI.}
\label{fig:braak_scatter}
\end{figure*}

Table~\ref{tab:braak_suvr} summarizes the SUVR-equivalent comparison between real and MCR-VQGAN-synthesized tau PET across Braak-defined ROIs. Strong linear associations were observed across all ROIs (Pearson $r = 0.78$--$0.88$) with Braak V/VI showing the highest correlation ($r = 0.877$). No significant mean-level difference was observed in Braak I/II ($p = 0.518$) and Other regions ($p = 0.156$), while synthetic values were slightly lower than real values in Braak III/IV ($p = 0.045$) and Braak V/VI ($p = 0.004$), reflecting mild underestimation in higher-tau regions. ICC values ranged from 0.708 (Other) to 0.838 (Braak V/VI), indicating moderate-to-good absolute agreement across all regions. Notably, the highest ICC was observed in Braak V/VI, the neocortical region most diagnostically informative for advanced AD. Fig.~\ref{fig:braak_scatter} presents scatter plots of real versus synthetic SUVR for each Braak ROI.

\begin{table*}[!t]
\caption{Regional SUVR comparison between real and MCR-VQGAN-synthesized tau PET images across Braak-defined ROIs ($n = 37$).}
\label{tab:braak_suvr}
\centering
\begin{tabular*}{\textwidth}{@{\extracolsep{\fill}}lcccccc@{}}
\toprule
{Braak ROI} & {Real SUVR} & {Synthetic SUVR} & {Mean Abs. Diff.} & {Pearson $r$} & {Paired $t$-test $p$} & {ICC} \\
\midrule
Braak I/II   & $1.111 \pm 0.169$ & $1.100 \pm 0.107$ & $0.0852 \pm 0.0668$ & 0.785 & 0.518 & 0.714 \\
Braak III/IV & $1.067 \pm 0.194$ & $1.026 \pm 0.131$ & $0.0835 \pm 0.0923$ & 0.806 & 0.045 & 0.732 \\
Braak V/VI   & $0.957 \pm 0.228$ & $0.901 \pm 0.192$ & $0.0864 \pm 0.0886$ & 0.877 & 0.004 & 0.838 \\
Other        & $1.012 \pm 0.129$ & $0.992 \pm 0.084$ & $0.0645 \pm 0.0549$ & 0.782 & 0.156 & 0.708 \\
\bottomrule
\end{tabular*}
\end{table*}

\section{Discussion}
In this study, we introduced MCR-VQGAN, a novel generative framework designed to synthesize high-fidelity tau PET images from structural T1-weighted MRI scans. By integrating multi-scale convolutions, ResNet blocks, and the Convolutional Block Attention Module (CBAM), MCR-VQGAN outperformed established GAN architectures in both quantitative reconstruction accuracy and visual fidelity. Our findings indicate that the proposed framework serves as a robust tool for cross-modality synthesis, providing a promising path toward a scalable surrogate for tau PET imaging.

The ablation study confirmed that each architectural innovation contributed incrementally to the model's superior performance. These consistent improvements are summarized in Table~\ref{tab:ablation} and illustrated in Fig.~\ref{fig:ablation_comp}. The most substantial gain came from the multi-scale convolutions, which expanded the receptive field to capture both local details and global anatomical context. This multi-scale approach is vital for mapping the complex, non-linear relationships between brain structure and tau deposition, aligning with previous evidence that multi-scale feature extraction is essential for high-fidelity medical image reconstruction~\cite{yuan2018multiscale,wang2023high}. Furthermore, ResNet blocks stabilized the adversarial training process and mitigated vanishing gradient issues~\cite{bengio1994learning,glorot2010understanding}, leading to better preservation of fine structural details. Lastly, CBAM enabled the model to selectively emphasize informative spatial and channel-wise features~\cite{woo2018cbam}. This synergy allowed for the preservation of fine structural details that are often lost or over-smoothed in standard VQGAN implementations.

The systematic hyperparameter optimization further validated the robustness of the proposed architecture. The codebook parameter evaluation revealed that all six configurations achieved comparable performance (SSIM range: $0.9228$--$0.9239$), indicating that MCR-VQGAN is not sensitive to codebook size selection. The generator loss weight grid search showed robust performance across a broad range of configurations (SSIM range: $0.9207$--$0.9255$), indicating that MCR-VQGAN is relatively insensitive to loss weight choice within this search space. The final configuration, $\lambda_\text{rec} = 10$, $\lambda_\text{perc} = 10$, $\lambda_\text{VQ} = 5$, was selected to balance reconstruction fidelity, perceptual quality, and codebook alignment, and demonstrated stable convergence under the full 500-epoch schedule.

In our comparative analysis, MCR-VQGAN consistently surpassed established generative models, including cGAN, CycleGAN, WGAN-GP, and standard VQGAN. As shown in Table~\ref{tab:comparison} and Fig.~\ref{fig:baseline_comp}, MCR-VQGAN achieved superior performance across all quantitative evaluation metrics and generated high-fidelity and structurally preserved tau PET images. In this study, MCR-VQGAN is the largest and most complex model, with approximately 75.0 million parameters and an inference time of 56.6 seconds. By comparison, the baseline models ranged from 4.2 to 22.7 million parameters with inference times between 38.7 and 42.4 seconds. Despite the substantial difference in parameter counts between MCR-VQGAN and the baseline models, the relatively small increase in inference time underscores the computational efficiency of MCR-VQGAN alongside its superior generative performance. Among the baseline models, WGAN-GP emerged as the strongest baseline model due to its stable training dynamics and resistance to model collapse in image synthesis~\cite{gulrajani2017improved}. In contrast, CycleGAN yielded the lowest performance. This outcome is expected given its reliance on unpaired image-to-image translation with cycle-consistency loss, which makes it less suitable for precise pixel-level mapping. 

As shown in Table~\ref{tab:cog_comparison} and Fig.~\ref{fig:group_comp}, our analysis of MCR-VQGAN's synthesis performance across different cognitive states revealed a critical observation. The fidelity of image synthesis decreased systematically as disease severity increased. This trend can be attributed to the clinical complexity of advanced AD. As AD progresses, the relationship between brain structure assessed through MRI and tau pathology visualized by PET becomes increasingly complex, unpredictable, and less spatially correlated due to severe atrophy and heterogeneous tau deposition patterns~\cite{jeon2019topographical,lu2023heterogeneity,rauchmann2025multimodal}. Additionally, the limited number of AD cases in the training set (n=9) likely hindered the model's ability to generalize to late-stage pathology. Expanding the cohort to include more advanced AD cases will be essential for improving the model's diagnostic reach.

The clinical utility of the synthetic images was evaluated in a downstream classification task to assess the preservation of pathological features. As summarized in Table~\ref{table_3}, the primary finding was that a classifier trained on real images achieved comparable performance when tested on both real (63.64\%) and synthetic (65.91\%) tau PET images. This suggests that the MCR-VQGAN synthesis preserves essential diagnostic features. Although the absolute classification accuracy is modest, our intent was to test for diagnostic information preservation rather than deployability. The comparable performance between real and synthetic tau PET indicates that MCR-VQGAN retains AD-relevant discriminative features.

The Braak staging framework analysis provides clinically grounded validation complementing pixel-level metrics. Strong linear correlations ($r = 0.78$--$0.88$) and moderate-to-good ICC values ($0.71$--$0.84$) indicate that MCR-VQGAN preserves Braak-regional tau patterns at both cohort and subject levels, with the strongest agreement observed in Braak V/VI ($r = 0.877$; ICC $= 0.838$). The slight underestimation in Braak III/IV and V/VI converges with the diagnostic-group analysis (Table~\ref{tab:cog_comparison} and Fig.~\ref{fig:group_comp}) in which fidelity decreased from CN to AD, most plausibly explained by the limited number of AD cases in the training set ($n = 9$). Because our pipeline operates in a normalized $[-1, 1]$ intensity space, our SUVR-equivalent values are not on the same numerical scale as conventional PET SUVRs, which precluded categorical Braak staging with literature-derived thresholds. Regional SUVR agreement therefore offers the most informative Braak staging framework evaluation accessible in this setting; extending the framework to absolute SUVR prediction is a promising direction for future work.

Despite these promising results, several limitations must be acknowledged. First, a major methodological limitation is that our model was trained and evaluated based on 2D images. While computationally efficient, this design does not leverage the full 3D spatial context of the volumetric data, which may restrict its ability to learn complex and non-linear pathological patterns. Second, the findings are based on a relatively small and imbalanced dataset. The limited number of subjects in the advanced AD cohort (n=3 in the test set) makes our subgroup analysis preliminary and prevents statistically robust conclusions on late-stage performance. The lack of an external dataset also means that the model's generalizability across different scanners and populations is unconfirmed. Third, while the downstream task suggests the preservation of some diagnostic information, the low absolute accuracy ($\sim$65\%) of the classifier itself means that these results do not support the immediate clinical utility of the synthetic images for standalone diagnosis. Finally, this study did not include an evaluation by clinical experts to assess the perceptual quality and diagnostic applicability of the synthetic images.

Future work should focus on expanding the dataset through additional ADNI data and other external data sources. Additionally, we can explore domain adaptation strategies to further improve the generalizability of our proposed method. Furthermore, we can extend our framework to a multimodal architecture that integrates other data modalities or longitudinal data, which are crucial in AD-related studies. Finally, future studies can improve the downstream task by conducting blinded expert-based evaluation studies, in which neuroradiologists will evaluate a balanced and randomly selected cohort across diagnostic classes (CN, MCI, AD). This blinded expert-based evaluation will provide stronger evidence of fidelity and clinical utility of images generated by the proposed model.

\section{Conclusion}
In this study, we presented MCR-VQGAN, a novel deep learning framework designed to synthesize high-fidelity tau PET images from structural T1-weighted MRI. By integrating multi-scale convolutions, ResNet blocks, and CBAM into the VQGAN architecture, our proposed model effectively captures both global anatomical context and pathological details. Our ablation study validated the incremental contribution of each architectural component. In addition, our comparison experimental results demonstrated that MCR-VQGAN outperforms established generative models, including cGAN, WGAN-GP, CycleGAN, and standard VQGAN, across quantitative metrics and visual quality. Furthermore, the downstream classification task indicated that the synthetic images preserve diagnostically relevant features, achieving classification accuracy comparable to real tau PET scans. These findings suggest that MCR-VQGAN has the potential to serve as a scalable, cost-effective, and non-invasive surrogate for conventional tau PET imaging. Future work will focus on validating the model on larger, multi-site external datasets and exploring multimodal integration to further enhance generalizability and clinical applicability in AD research and clinical workflows.



\bibliographystyle{IEEEtran}
\bibliography{references}
\end{document}